# Sigma meson resonance and its coupling constant

M. C. Menchaca Maciel and J. R. Morones Ibarra

Facultad De Ciencias Físico-Matemáticas, Universidad Autónoma de Nuevo León,
Ciudad Universitaria, San Nicolás de los Garza, Nuevo León, 66450, México.

We study two coupling constants reported as predictions of the sigma model. We obtain an spectral distribution, in terms of the σ meson regularized self- energy in one loop order, which is consistently with the Breit-Wigner distribution, converging to values of mass and width of this scalar resonance given in the experimental results reported by the E791 Collaboration at Fermilab and BES Collaboration at the Beijing Electron Positron Collider. We conclude that the sigma meson mass is restricted to be on the range 453MeV $< m_\sigma <$ 490MeV and its width 238MeV-354MeV or 354MeV $< m_\sigma <$ 415MeV and its width 167.5 $< \Gamma_\sigma <$ 347MeV MeV.

## 1. INTRODUCTION

Recent experimental results such as $D^+ \to \sigma\pi^+ \to \pi^+\pi^+\pi^-$ [1], and $J/\psi \to \omega\pi^+\pi^-$ [2] supports the existence of σ meson. Several treatments have been proposed to explain the properties of this particle, the dispersion analysis employing the Roy equation and the Dalitz Plot are part of these intensive researches. However, the existence or non-existence of the sigma meson is still a source of debates due to the role it plays in nuclear and hadronic physics [3].

The mass and width of resonance can be define in three different ways: Lagrangian mass parameters, Breit-Wigner mass and Second sheet pole position (T-matrix) [4].

Still when some experts say that the Breit-Wigner form does not really give a satisfactory description of the scalar resonance shape, it certainly is an important definition of the mass because the results of the E791 Collaboration at Fermilab belongs to this category [5].

The use of an "junk entry" [6], that is the idea that we can use any entry as the sigma meson mass [7], maybe arise from the modified propagator obtained via the twice subtraction and the definition of the observable mass:

$$\text{Re}[\Delta(k^2)^{-1}] = k^2 - m_\sigma^2 - \text{Re}\Sigma^R(k^2) + \text{Re}\Sigma^R(m_\sigma^2) = 0 \qquad (1)$$

From Eq.(1) seems that all values k=$m_\sigma$, can be used. However we show that the self-intersection point in the graph of Eq.(1) is in fact the maximum of an harmonic system, and the renormalization of the bare mass can be used to understand the resonance and not only as a method to eliminate the infinite terms of ReΣ($k^2$). The value of $m_\sigma$ gets restricted to a very narrow range, if we demand a well behavior of $\text{Re}[\Delta(k^2)^{-1}]$.

In this work, we focus on the spectral function and the reasons of why this can fail to reproduce the experimental results for the sigma meson. The main technical problems that we would have solved in this work are: the use of the "junk entry", the differences between the observable mass and the peak of the resonance and finally the calculation of the width.

All of these issues can be solved by use of the suitable coupling constant and the original definitions arising from the Breit-Wigner distribution and the linear sigma model.

The organization of this paper is as follows. In Sec. 2, we explain the basic concepts and illustrate how we can establish a first approximation for the physical mass. The importance of the behavior real part of the inverse propagator is emphasized and a second approximation for the mass are illustrated in Sec. 3. In Sec. 4, we discuss why the peak of the resonance given by the typical spectral function does not coincide with the observable mass and we propose a new spectral distribution. In Sec 5, we discuss about the BES experiment and the coupling constant arising from the global color symmetry model.

## 2. MODIFICATION OF σ-MESON PROPAGATOR

The modified σ -meson propagator is obtained through the modification of the free σ -meson propagator in the one loop approximation. The full propagator can be calculated with the chain approximation method.

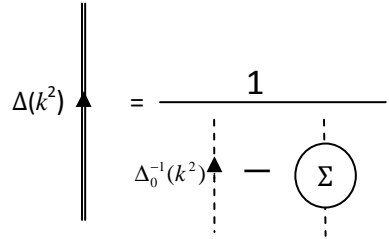

**Figure 1.** Sum over all repetitions of all self-energy parts of any number of pion "bubbles" inserted into the bar σ-meson propagator $\Delta_0(k^2)$

As indicated in Figure1, the dressed or modified σ – meson propagator [8] obtained from the Dyson equation is

$$\Delta(k^2) = \frac{1}{\Delta_0^{-1}(k^2) - \Sigma(k^2)} \quad (2)$$

where

$$\Delta_0^{-1}(k^2) = k^2 - (m_\sigma^0)^2 \quad (3)$$

$m_\sigma^0$ is the bare mass of the σ -meson and the self-energy $\Sigma(k^2)$ needs to be calculated.

The simplest renormalizable π-σ coupling obtained from the σ model is given by

$$L_{\sigma\pi\pi}^{eff} = \frac{1}{2} g_{\sigma\pi\pi} m_\pi \vec{\pi} \cdot \vec{\pi} \sigma \quad (4)$$

The coupling constant [9] $g_{\sigma\pi\pi}$ becomes from the relation of the σ model between the masses and the coupling constant of σ and the ππ system. In our scheme $g_{\sigma\pi\pi}$ is dimensionless and real:

$$g_{\sigma\pi\pi} = \frac{m_\sigma^2 - m_\pi^2}{2 f_\pi^2} \quad (5)$$

$f_\pi$ =92.4MeV [10] is the pion decay constant and $m_\pi$ =139.57MeV the pion mass [11].

The self-energy has the analytical expression given by the two-pion-bubble integral

$$-i\Sigma(k^2) = \frac{3 g_{\sigma\pi\pi}^2 m_\pi^2}{2} \int \frac{d^4 q}{(2\pi)^4} \cdot \frac{1}{q^2 - m_\pi^2} \cdot \frac{1}{(q-k)^2 - m_\pi^2} \quad (6)$$

The constants are suitable for $\vec{\pi} \cdot \vec{\pi}\sigma$ coupling of charge-symmetric pions to the σ meson [12].
The final results for the self energy:

$$\Sigma^R(k^2) = \frac{g_{\sigma\pi\pi}^2}{64\pi^2 \sqrt{k^2(k^2 - 4m_\pi^2)}} [(k^2 - 12m_\pi^2)\sqrt{k^2(k^2 - 4m_\pi^2)} + 6m_\pi^2(k^2 - 4m_\pi^2)\{Ln\left|\frac{\sqrt{k^2(k^2 - 4m_\pi^2)} + k^2}{2m_\pi^2} - 1\right| + \pi i\}] \quad (7)$$

The divergence of the integral for the real part will be absorbed into the renormalization of the bare mass $m_\sigma^0$, the renormalized mass may be set equal to this physical mass by appropriate choice of counter terms:

$$m_\sigma^2 = (m_\sigma^0)^2 + \text{Re}\,\Sigma(m_\sigma^2) \tag{8}$$

the modified propagator takes the form

$$\Delta(k^2) = \frac{1}{k^2 - m_\sigma^2 - \Sigma^R(k^2) + \text{Re}\,\Sigma^R(m_\sigma^2)} \tag{9}$$

If Eq.(1) is satisfied, Eq. (5) can be rewritten as

$$g_{\sigma\pi\pi} = \frac{k^2 - m_\pi^2}{2 f_\pi^2}\bigg|_{k^2 = m_\sigma^2} \tag{10}$$

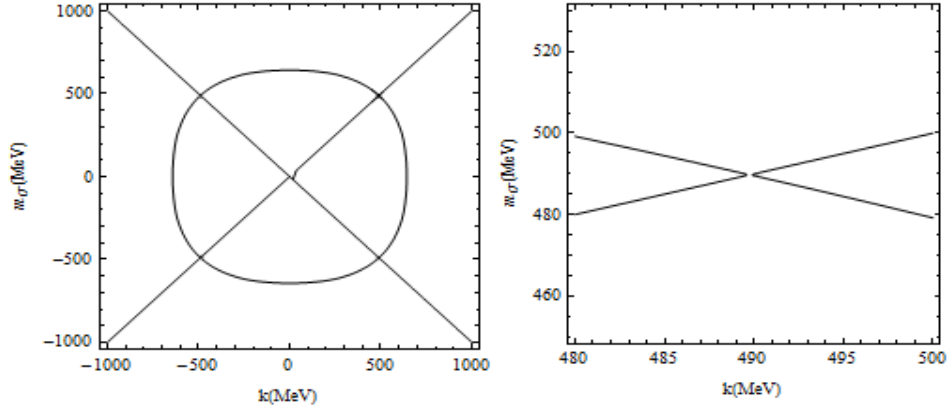

**Figure 2.** Contour Plot for $\text{Re}[\Delta(k^2)^{-1}] = 0$. The left panel shows the self-intersection point at $m_\sigma = k \approx 489.8$ MeV. The right panel shows that there exist a "jump" near to the self-intersection point. This jump suggest that the resonance takes place near to this value.

Let us considered the case where we do not know the quantity that cancel the infinite terms of Re$\Sigma(k^2)$, that is, the renormalized mass is merely a bookkeeping device that has no physical content.
The modified σ – meson propagator becomes

$$\Delta(k^2, m_R^2) = \frac{1}{k^2 - m_R^2 - \Sigma^R(k^2)} \tag{11}$$

The contour plot of $\text{Re}[\Delta(k^2, m_R^2)^{-1}] = 0$, is an harmonic system, when $k = m_\sigma$ and $m_R$ are not restricted.

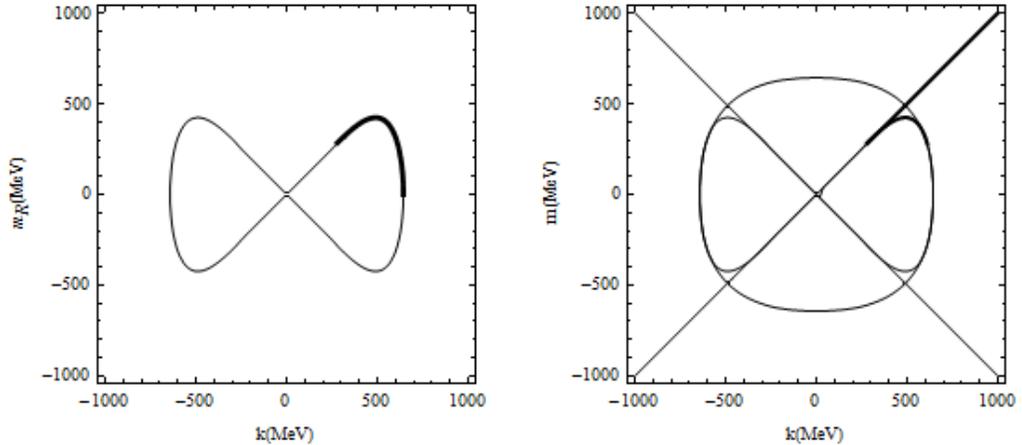

**Figure 3.** Contour Plot for $\text{Re}[\Delta(k^2, m_R^2)^{-1}] = 0$. Left panel: the curve represents the real part of the inverse propagator equal zero and the set of solutions when $k = m_\sigma > 2m_\pi$ and $m_R > 0$ has been thickened. Right panel: the real part of the inverse propagator equal zero using the twice subtraction and the renormalized mass in the same screen. The set of solutions with the restriction for the masses has been thickened.

The maximum of the Figure3 in the left panel is at $m_\sigma = k = 489.761$ MeV. For $k = m_\sigma > 2m_\pi$ and $m_R > 0$. We can approximate $\text{Re}[\Delta(k, m_R)^{-1}] = 0$ with the parametric equations:

$$m_\sigma = 642.980 \, MeV \, \cos[t - \frac{\pi}{40}],$$
$$m_R = 423.554 \, MeV \, \sin[2t], \tag{12}$$
$$0 \leq t \leq 1.2$$

and the width of the spectral distribution can be related with the parameter t as

$$\Gamma \approx \frac{m_\sigma}{t} - \Lambda \tag{13}$$

Where $\Lambda = 285$ MeV.

## 3. THE REAL PART OF THE INVERSE PROPAGATOR

The observable mass of $\sigma$ is defined as a zero of the real part of the inverse propagator, which must be not differentiable at $k = 2m_\pi$ and its solution unique. From the parametric equations we conclude that $2m_\pi < m_\sigma < 643$ MeV, however we need to investigate the graph of $\text{Re}[\Delta(k^2)^{-1}]$. For $m_\sigma > 560$ MeV the graph cross the k-axis three times, therefore $m_\sigma \leq 560$ MeV. $\text{Re}[\Delta(k^2, m_\sigma^2)^{-1}] > \text{Re}[\Delta((k^2, 453.39 MeV)^2)^{-1}]$ at $k \to 2m_\pi$. The value $m_\sigma = 453.39$ MeV seems to be the minimum value that we can take in order to obtain a well behavior of the real part of the inverse propagator.

Our preliminary results agree with quark model that predict that the mass cannot be as low as 390 MeV [13] and the experimental value $m_\sigma = 483 \pm 31$ MeV.

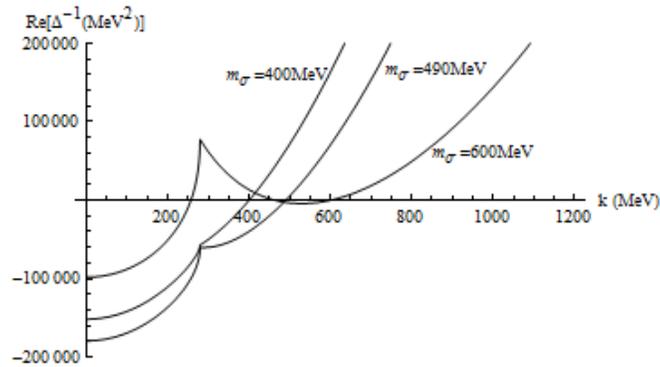

FIGURE 4  Graph of $\text{Re}[\Delta(k^2)^{-1}]$ for different values of the meson mass.

## 4. THE SPECTRAL FUNCTION AND THE BREIT-WIGNER DISTRIBUTION

The spectral function can be expressed in terms of the retarded self-energy as

$$S(k^2) = \frac{-\operatorname{Im}\Sigma(k^2)}{\left[k^2 - m_\sigma^2 + \operatorname{Re}\Sigma^R(m_\sigma^2) - \operatorname{Re}\Sigma^R(k^2)\right]^2 + \left[\operatorname{Im}\Sigma(k^2)\right]^2} \quad (14)$$

The normalization condition

$$\int_0^\infty S \, dk = 1 \quad (15)$$

does not holds for Eq. (14), the spectral function has to be normalized by hand.

$$S(k^2) = \frac{\phi_s \operatorname{Im}\Sigma(k^2)}{\left[k^2 - m_\sigma^2 + \operatorname{Re}\Sigma^R(m_\sigma^2) - \operatorname{Re}\Sigma^R(k^2)\right]^2 + \left[\operatorname{Im}\Sigma(k^2)\right]^2} \quad (16)$$

The peak of the spectral function is always at the right of the value used as BW mass. The spectral function is asymmetric and the kurtosis can be positive or negative for some values of the mass. For $m_\sigma < 453.39$ MeV, the spectral function has positive kurtosis and for $m_\sigma > 489.761$ MeV, the spectral function has negative kurtosis.

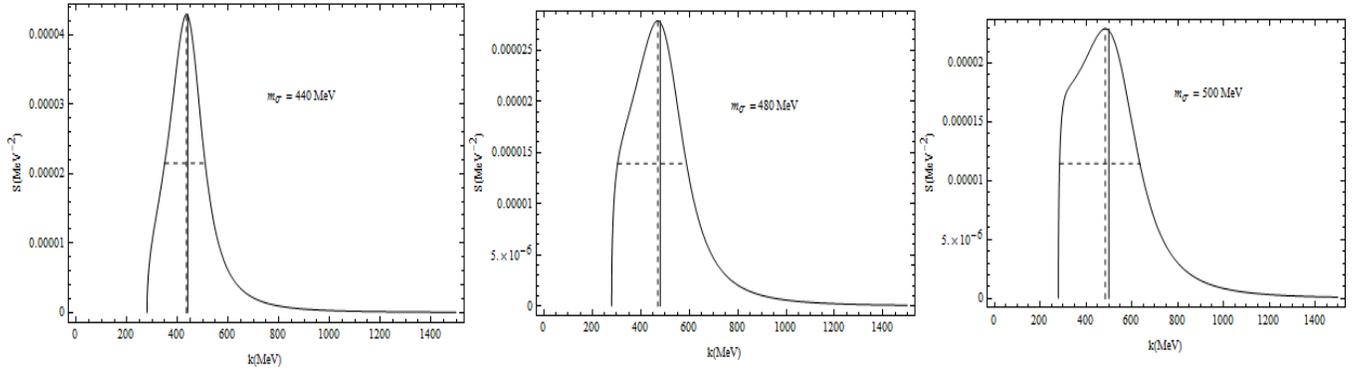

**Figure 5.** Spectral function for different values of de mass. The width of the spectral function is taken at one half of the maximum value of k (dashed vertical line) and the sigma meson mass (continuous vertical line) is taken as the zero of the real part of the inverse propagator.

From this analysis, the spectral function is most likely to be considered as a mass distribution when the mass is restricted to be on the range 453MeV $< m_\sigma <$ 490MeV. Therefore the calculations with the coupling constant used in Section II. can be used for the Fermilab experiment but not for the BES experiment. Also we can compare our result with Dalitz plot where the scalar resonance was determined with 478MeV.

The original Breit-Wigner distribution has the form:

$$SBW(k; m_\sigma, \frac{\Gamma}{2}, 1) = \frac{\left(\frac{\Gamma}{2}\right)^2}{[k - m_\sigma]^2 + \left(\frac{\Gamma}{2}\right)^2} \quad (17)$$

We have a three-parameter Lorentzian function where the height of the peak is 1.

Using z scores for the k parameter we get the standard Cauchy(0,1) distribution. Notice that the distribution of z scores has the same shape as the original distribution, therefore they are *not* a remedy to obtain a normal distribution data, important assumption in the Cauchy distribution.

We modified the spectral function in order to obtain the spectral distribution M(k), that is consistently with the peak of the Breit-Wigner distribution

$$M(k) = \frac{\left[\text{Im}\Sigma(k^2)\right]^2}{2k^2\left[k - \sqrt{m_\sigma^2 + \text{Re}\Sigma^R(m_\sigma^2) - \text{Re}\Sigma^R(k^2)}\right]^2 + \left[\text{Im}\Sigma(k^2)\right]^2} \tag{18}$$

Let us consider the central values of the E791 Collaboration and Dalitz plot to find the width of the resonance using the quantum spectral function given by Eq.(16) and comparing the result with Eq.(18). For the spectral function and spectral distribution the width is taken at one half of the maximum value and the mass as the peak of the resonance. For a mass of 483MeV, the coupling constant of the strong decay σ→ππ is 12.5214.

**Table 1.** Experimental results from E791 Collaboration against theoretical results of $S(k^2)$ and M(k).

|  | $m_\sigma$ (MeV) | Γ (MeV) |
|---|---|---|
| Fermilab | 483 | 338 |
| $S(k^2)$ | 472.263 | 295.195 |
| M(k) | 483 | 330.469 |

We used the t-test procedure to test whether the experimental value differs from that in theoretical values. The test indicate that the values obtained from the quantum spectral function differs significantly from the experimental values of Fermilab and Dalitz plot, while the values obtained from the spectral distribution do not differs significantly from the experimental values of Fermilab and Dalitz plot.

**Table 2.** Amplitude analysis of the 3π Dalitz plot against theoretical results of $S(k^2)$ and M(k).

|  | $m_\sigma$ (MeV) | Γ (MeV) |
|---|---|---|
| Dalitz plot | 478 | 324 |
| $S(k^2)$ | 468.301 | 276.518 |
| M(k) | 478 | 314.094 |

## 5. THE COUPLING CONSTANT IN THE GLOBAL COLOR SYMMETRY MODEL AND THE J/ψ DECAYS

The coupling constant in Eq.(5) arise from the ordinary linear sigma model, in the global color symmetry model [14] the coupling constant was calculated as

$$f_{\sigma\pi\pi} = \frac{m_\sigma^2}{f_\pi} \quad (19)$$

This coupling constant was used as the prediction of the linear sigma model.
In our scheme this equation as the equivalence

$$g_{\sigma\pi\pi} = \left.\frac{k^2}{m_\pi f_\pi}\right]_{k^2 = m_\sigma^2} \quad (20)$$

Substituting this coupling constant in the renormalized self-energy to obtain the mass for the sigma meson through analysis of the inverse propagator we get that the self-intersection point is at $m_\sigma = k \approx 414.9$ MeV.

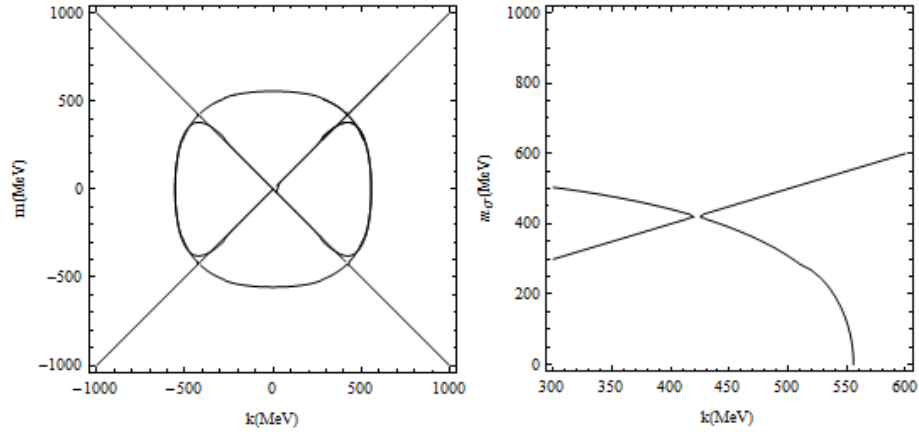

**Figure 6.** Contour Plot for $\mathrm{Re}[\Delta(k^2)^{-1}] = 0$. The left panel: the real part of the inverse propagator equal zero using the twice subtraction and the renormalized mass in the same screen. The right panel shows that there exist a "jump" near to the self-intersection point $m_\sigma = k \approx 414.9$ MeV.

As preliminary study, the sigma meson resonance takes place near to the value of 415MeV, also we have that $m_\sigma \leq$ 560 MeV. We can find the width with the quantum spectral function and the spectral distribution using the central value of the BES Collaboration[15,16] where the coupling constant of the strong decay σ→ππ is 11.7941.
The width obtained with the new distribution is a better approximation than the value obtained with the quantum spectral function.

Table 3. Experimental results from BES Collaboration against theoretical results of $S(k^2)$ and $M(k)$.

|        | $m_\sigma$ (MeV) | Γ (MeV) |
|--------|------------------|---------|
| BES    | 390              | 282     |
| $S(k^2)$ | 318            | 206.048 |
| M(k)   | 390              | 263.968 |

The graph of $\text{Re}[\Delta(k^2)^{-1}]$ cross the k-axis more than once for any value of $m_\sigma > 450$. Therefore, we conclude that the sigma meson is restricted to be on the range 354MeV $< m_\sigma <$ 415MeV and its width $167.5 < \Gamma_\sigma <$ 347MeV MeV.

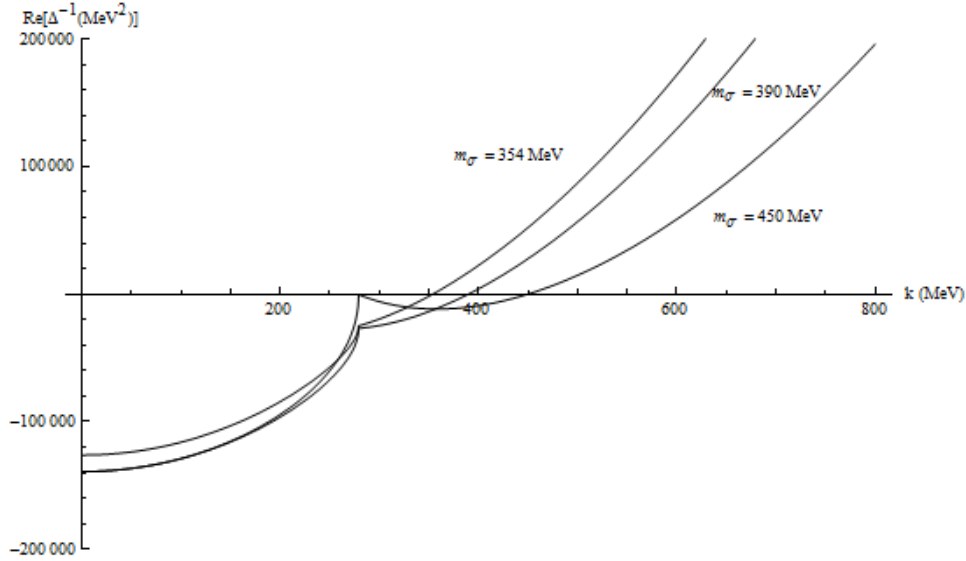

**Figure 7.** Graph of $\text{Re}[\Delta(k^2)^{-1}]$ for different values of the meson mass.

## 6. CONCLUSIONS

The nature of the $f_0(600)$ or sigma meson is far from being resolved, however we found that if the sigma meson can be considered as two pion resonance the mass is restricted to be on the range 453MeV $< m_\sigma <$ 490MeV and its width 238-354MeV when the coupling constant arising from the sigma model is used. By the other hand, the coupling constant for the BES experiment give us the range 354MeV $< m_\sigma <$ 415MeV and its width $167.5 < \Gamma_\sigma <$ 347MeV MeV. The difference is not even close to being statistically significant for the mass and width given by the Spectral distribution and the Fermilab or BES experiment. Also the maximum in the harmonic system seems to be the value where the resonance takes place and correspond to the maximum value on the range.

**ACKNOWLEDGEMENTS**
The authors gratefully acknowledge PhD Scholarship Programmer from CONACYT